\documentclass[11pt, a4paper]{article}
\usepackage[utf8]{inputenc}
\usepackage[margin=33mm]{geometry}
\usepackage{graphicx}
\usepackage{authblk}
\usepackage{amsmath}
\usepackage{natbib} 
\setcitestyle{authoryear,open={(},close={)}} 

\title{The scalar mismatch of regional governance: a comparative analysis of hierarchical structures}
\author[1,*]{Valentina Marin}
\author[1]{Carlos Molinero}
\author[1]{Elsa Arcaute}
\affil[1]{Centre for Advanced Spatial Analysis (CASA) University College London, UK}
\affil[*]{corresponding author: v.marin@ucl.ac.uk}
\date{}

\begin{document}
\maketitle

\begin{abstract}
Self-organisation in territories leads to the emergence of patterns in urban systems that shape the interactions and dependencies between cities, resulting in a hierarchical organisation. Governance follows as well a hierarchical structure, breaking the territory into smaller units for its management. The possible mismatch between this two different types of organisations may lead to a range of problems, ranging from inefficiencies to insufficient and uneven distribution of resources. 
This paper seeks to develop a methodology to explore and quantify the correspondence between the hierarchical organisation given by the structure of governance and that given by the structure of the urban systems being governed, where Chile is used as a case study. 
The urban hierarchical structure is defined according to the connectivity of the system given by the road network. This is extracted through a clustering algorithm defined as a percolation process on the Chilean street network, giving rise to urban clusters at different scales. These are then compared to the spatial scales of the politico-administrative system.
This is achieved by using measures of pair-wise distance similarity on the dendrograms, such as the cophenetic distance, by looking at the different clustering membership using Jaccard similarity, and by analysing the topological diversity defined as the structural entropy. 
The results show that the urban sub-national structures present high heterogeneity, while the administrative system is highly homogeneous, replicating the same structure of organisation across the national territory. 
Such contrasting organisational structures present administrative challenges that can give rise to poor decision-making processes and mismanagement, in addition to impairing the efficient functioning of the systems themselves. 
Our results can help address these challenges, informing how to rebalance such mismatches through planning and political strategies that consider the complex interdependencies of territories across scales.


\end{abstract}

\section{Introduction}

Politics are inherently geographical and scalar \citep{Agnew1997}. Space is an important element of governance because governmental activities are territorially demarcated and have well-defined scales of operation \citep{Hudson2007}. These scales are spatially defined, so the territory over which governance is exercised is delimited by clear geographical boundaries such as communes, provinces and regions. In some cases, these boundaries are based on traditional and historical identities, while in others, they are imposed for utilitarian purposes and have little relation to existing territorial dynamics \citep{Terlouw2012}. The spatial extension of different levels of governance is often defined by the central state to accommodate administrative functions. State institutions play a crucial role in creating, modifying or removing boundaries and defining scalar organisations and partitions. In centralised governments, decision-making processes tend to occur distant to regional and local spaces of governance, leading to the implementation of standardised policies that do not account for the local particularities. In contrast, polycentric structures of governance operate at multiple scales, allowing for policies that better match territorial diversity \citep{Lebel2006, Cumming2013}. Developing effective regional policy requires relevant regional scalar data and a thorough understanding of regional structures. This information is key for decision-makers to identify the unique needs and challenges of different regions, allowing them to tailor their policies and plans accordingly \citep{Rozenblat2009}.

\subsection*{Scales of management and the scales of processes}

The scalar thinking permeates different layers of organisation: the political, the geographical, the economical, the sociological  and the infrastructural. 
The ubiquitous use of the concept has raised concern among scholars who stress the tension between different notions and conflated conceptions of scale \citep {Smith1992, jonas1994, Marston2000, Brenner2001, Moore2008, MacKinnon2011, Marston2016}. \citet{jonas1994} and \citet{Moore2008} have underlined the distinction between ``operational" and ``discursive" approaches to the scalar thinking. The operational scale is considered the real scale of processes where the sociospatial interactions occur. Discursive scales on the other hand, are a frame or set of abstractions used to understand and study the real processes \citep{jonas1994, Moore2008,Lukas2019}. In this context, discursive scales are the outcome of imposed boundaries \citep{Lukas2019}, which are used to identify phenomena within a system of interest by setting "research scales", or to implement policies relevant to the phenomena through "management scales". The definition of discursive scales has long been a subject of controversy, as they can be influenced by embedded assumptions, political implications, and practicality considerations \citep{Rangan2009}. Some scalar narratives have been widely accepted without adequate scientific support, leading to the production of knowledge that is inconsistent with real processes \citep{Lukas2019}. Geographical data for example, is often provided aggregated into spatial jurisdictions and not necessarily on the basis of ``scientifically-derived" scales \citep{jonas1994}. Administrative boundaries or jurisdictions have been used for a long time to address urban issues either by governance, research, planning, or design approaches. These pre-given limits, provide all-purpose and durable in time urban definitions that can be easily implemented for a wide range of policies, plans and projects. Despite how convenient the straightforward use of these definitions may be, they are not always capable of reflecting the extent of the urban processes in question. Along with this, the speed of urban growth, the fuzzy urban-rural transition, and the multiple extent of economic, ecological, social and political processes and their different scales, are challenging this fixed and ``one-size-fits-all" approach \citep{Rozenfeld2008}. According to Moore (2008), the attempt to fit complex processes into reduced pre-given scales ends up overlooking the diversity of processes, ignoring the granularity and variability of sociospatial systems \citep{Moore2008}.

\subsection*{Scalar mismatch}

According to Cumming (2006), scalar mismatches take place when the scale of management and the scale of processes are aligned in such a way that inefficiencies occur either affecting the functioning of the systems being managed or the functioning of the institutions responsible for their management \citep{Cumming2013}. Governance in these cases is less sensitive to the variety of local structures, lacking the appropriate information and oversimplifying the knowledge about processes. Deficient monitoring frameworks, together with the lack of clear responsibilities and capability necessary to achieve the required scale of management, results in the formulation of misleading and inconsistent policies \citep{Cumming2006}. Under these circumstances, self-maintenance and self-organisation are undermined, affecting not only daily operations, but leading to a potential decrease of the urban resilience capacity \citep{Folke2005,Cumming2006}. Redressing scalar mismatches is therefore a crucial requirement to redefine the relationship between governance, planning, and sociospatial processes. 
Alternative geographies of governance more responsive to scalar processes as \citet{Bulkeley2005} suggests, need to be articulated.
Managing scalar mismatches requires that scale is not taken for granted. Therefore, the organisational structure of systems needs to be identified beyond pre-given administrative definitions \citep{Bulkeley2005}. In order \textit{``to govern, it is necessary to render visible the space over which government is to be exercised."} \citep[p.~36]{rose_1999}. The diversity and specificities of the territory need to be decoded, so that political decisions and planning strategies are not blind to the scalar particularities and complexities of the urban organisation. This challenges the understanding of sub-national scales such as communes, provinces or regions, as self-enclosed homogeneous entities and coherent bounded territories \citep{Brenner2001,Hudson2007}. Rather, it puts forward its conceptualisation as relational entities made up of intricate systems of cities.

\subsection*{Decoding and comparing hierarchical structures}

Urban systems are inherently relational \citep{Pumain2006i}, therefore, the identification of different intensities of connectivity patterns is a key criterion for their definition \citep{Bretagnolle2006, Pumain2006}. As objects of study, systems of cities could be defined by the nature of relationships holding them together \citep{batty2018inventing, Rozenblat2022}. Sub-national scales are then, the result of constituent elements and the way in which they are progressively connected to all others \citep{BattyFractal}. By ordering connections or similarities at successive scales according to their orders of magnitude, a hierarchical structure can be discerned. Subsystems consist of strong internal connections, while inter-systems present weaker ties \citep{Simon1995}. As \citet{Simon1977} observed \textit{``everything is connected, but somethings are more connected than others"}. And it is from this process of heterogeneous interactions that the hierarchical structure of the urban territory emerges from the most local scale to the largest. 

Hierarchical structures can be represented through dendrograms, which are branching graphical representations of hierarchical organisations \citep{Fowlkes1983}. To compare different dendrograms, certain structural similarities and differences can be visualised and quantified. The first methods were developed in biology for phenetic studies to evaluate results obtained from numerical taxonomic analysis. Distances on the basis of similarity between dendrogram elements, are widely used methods ever since \citet{Sokal1962} introduced the use of cophenetic distance. Although its use is still more prominent in disciplines such as biology and ecology, it has been extended to other fields to assess the efficiency of different clustering methods and the consistency between outputs \citep{Fischer1980,Rohlf1982,Saracli2013}. 

In this paper, we compare hierarchical structures or dendrograms in order to explore scalar mismatches. Specifically, we examine the dendrogram of management scales resulting from the hierarchical structure of the administrative system, and the dendrogram of operational scales, resulting from the percolation analysis given the connectivity of the urban layout. Using Chile as a case study, we compare both dendrograms via three methods. In the first method, we conduct a pairwise comparison of the hierarchical distance of each city to all other cities in the system using the cophenetic distance. This provides information on the variation in relationship intensity between systems of cities in both structures. Then we compare the clusters in terms of their membership using Jaccard similarity to quantify how closely the city systems are grouped together in accordance with administrative boundary definitions. Finally, we examine the topological structure of the dendrograms by measuring the diversity of the furcations using structural entropy. 
The results of these analyses reveal the centralised organisation of the national system, the heterogeneous intra-regional structures, and the relationships between systems of cities that go beyond administrative definitions.

\section{Case study}

\subsection*{Chilean decentralisation, a challenge for regional governance}

The current politico-administrative division of Chile dates back to 1974, when it was established during the early years of the military dictatorship. The regionalisation process carried out at that time was seen as a political control strategy managed by the central government to strengthen national security, exercise administration, and ensure the efficiency of the state \citep{Arenasetal2007}. The division of the territory into thirteen ``homogeneous and equivalent" regions \citep{CONARA}, was more about distributing power across the country than about empowering self-governing regions  \citep{OECD2017}. As a result, the spatial structure of the governance system in Chile was primarily based on utilitarian purposes, ignoring territorial identities and the intrinsic local organisation. The regional administrative structure is homogeneously repeated throughout the country and does not necessarily align with the scalar logic of urban processes. Each region is organised mono-centrally around a capital city, which serves as the hub of development and main source of services. The regional capital is supposed to host the main centres of employment, education, and services, as well as the government's departments and sub-national authorities. This political organisation has reinforced intra-urban regional disparities. In response to increasing demands for local identity and representation, three regions were re-delimited to better match the logic of local scales (Tarapacá and Los Lagos in 2007, and Bíobio in 2017), while the rest of the territorial jurisdictions remained largely unchanged. The current Chilean politico-administrative system is organised into a simple 4-tier structure consisting of communes (346), provinces (56), regions (16), and the national level. The structure is equivalent at any jurisdictional level, with non-intersecting memberships, and does not consider intermediate scales or temporary jurisdictions for specific policy purposes.

Chile's regional level is the core of its decentralisation policies and reforms, and is considered a crucial link between the national and local levels. Despite various attempts by democratic governments to promote regional reforms and different changes to the electoral system, Chile remains one of the most centralised countries in the OECD \citep{OECD2017}. The urgent need for a governance perspective that achieves the definitive decentralisation of the country has positioned regional governance as a top priority. To this end, the paper will focus more closely on the multiscalar organisation of the regions.

\section{Methods}

The methodology of the paper is divided into two main sections. The first section outlines how to extract the hierarchical organisation of the system through percolation analysis, and then how to represent it in terms of a dendrogram. The second section describes the various methods used to compare the different dendrograms and evaluate the potential mismatches between them. These are the cophenetic distance, the Jaccard similarity and the structural entropy. These steps will allow us to analyse the correspondence between the politico-administrative system and the urban organisation emerging from the country's inner structural connectivity.

\subsection*{Hierarchical organisation}

We use percolation analysis to retrieve the scalar structure of Chile based on the connectivity of the street network. Cultural, political and socio-historical processes have left traces in the way local entities organise. As explained in \citet{Arcaute2016}, these footprints are contained in the street network, representing the main proxy for communication and exchange between settlements. The street network plays a crucial role in the organisation of urban areas, serving as the setting for urban life and the location for many social and economic interactions \citep{Marshall2018}. Streets form the underlying network for the functioning of urban systems, enabling the reachability of services and facilities and the flow of people and goods between all elements of the urban system \citep{Strano2012}. The efficiency of these interactions is closely tied to the topological properties of the street pattern given by its connectivity \citep{Cavallaro2014}. Junctions, or intersections, play a key role in ensuring connectivity and providing access to a range of alternative routes. The more junctions there are, the greater the permeability of the network, the denser the urban activities, and the greater potential there is for urban interaction. 

\textbf{Percolation analysis:} Percolation theory can be applied to identify substructures in a system describing phase transitions of connected clusters that characterise the potential flow of information within the structure \citep{Gallos2011}. It provides information on the global scale structure of connectivity of a system from the bottom-up. The process starts with disconnected elements that merge together based on a distance threshold parameter until the largest cluster covers the entire system. Percolation-like processes have been applied to different fields of research like functional brain networks \citep{Gallos2011}, the spread of diseases \citep{Newman2002, Gallos2012} and indeed, they have been used to study the structure of urban systems \citep{Rozenfeld2008,Rozenfeld2011,Fluschnik2016,Behnisch2019,Piovani2017}. In \citep{Arcaute2016}, percolation was used to uncover the urban hierarchical organisation of Britain through multiple percolation transitions based on the distance between street intersections.

This research applies percolation processes to the Chilean infrastructural network to uncover its hierarchical structure and identify the relevant scales of analysis. In this work we follow the procedure used in \citep{Arcaute2016,Piovani2017}. The street network is obtained from OpenStreetMap (OSM) as the primary source of our spatial data. The OSM data contains detailed information on the street network, including attributes such as speed limits. The network is represented by nodes (intersections) and links (street segments between intersections) that are embedded in space and weighted by the time required to travel by car along each link. The decision to weight the links of the network using time, allows us to include ferries' routes, which are the only means of transportation in the southern archipelagos. This approach accounts for the travel time cost and waiting times associated with these routes, allowing for their effective inclusion in the analysis. Using time as the weighting attribute provides a more comprehensive representation of the movement between spatial nodes. In this way, we are able to capture the main structure of connectivity throughout the territory. The process consists in extracting the sub-graphs of connected links whose weight $w_{ij}$ is lower than a given threshold $\tau$, $w_{ij} \leq \tau$. The process starts from the minimum time threshold at which all links are disconnected. Increasing the time threshold, leads to a series of percolation transitions where connected links with $w_{ij} \leq \tau$ form clusters. The clusters appearing at lower thresholds, i.e. for smaller travel time, do not necessarily imply physical proximity, but rather a greater density of intersections, and hence a higher potential for urban interactions. Important scale transitions can be identified when discontinuities occur in the evolution of the largest cluster size across the different time thresholds \citep{Arcaute2016}. These discontinuities do not imply actual breaks in the structure, but rather changes in the intensity and connectedness of the interactions  \citep{Simon1962,Holling2001UnderstandingSystems}. Thresholds where these discontinuities occur are selected as the main levels of the hierarchical structure.  As the threshold increases, cities are captured first, followed by urban conurbations, systems of cities and regions, until the entire country is obtained at the end of the process.

\textbf{Dendrogram visualisation:} The hierarchical structure of the urban system is reconstructed by ordering the different clusters obtained through the above-mentioned thresholding procedure. This is done using a dendrogram or tree-like structure that is organised vertically into different levels or scales, each representing a selected threshold with the highest at the top and the lowest at the bottom. This is possible because clusters generated using a lower threshold $\tau_1$ are contained within the clusters obtained using a higher threshold $\tau_2$. Horizontally, clusters have strong intra-cluster interactions and weaker inter-cluster linkages. At the same hierarchical level, clusters are not closely connected and instead communicate at higher scales through cross-scalar relationships. The dendrogram shows the largest cluster of the system (in this case, the entire country) at the top, branching down and decomposing into subsystems according to the strength of network connectivity, until reaching the terminal nodes at the bottom scale (cities and towns). On the other hand, the dendrogram representing the politico-administrative structure is constructed based on the nested territorial subdivision of communes, provinces, and regions, which are taken as hierarchical scales. To facilitate readability, we will refer to the governance dendrogram as $\varphi_g$ and to the dendrogram of the urban structure resulting from the percolation as $\varphi_s$. Both dendrograms have the same number of terminal clusters and use the same colour coding to indicate regional membership, which allows for easier comparison. For each dendrogram let $N=\{n_1, n_2, n_3, ...,n_k\}$ be the terminal clusters capturing cities and towns and $C_\tau=\{c_1,c_2,c_3, ...,c_k\}$ all the clusters at a specific threshold $\tau$. 

\subsection*{Hierarchical comparison}

\begin{figure}[h!]
\centering
\makebox[\textwidth][c]{\includegraphics[width=1\textwidth]{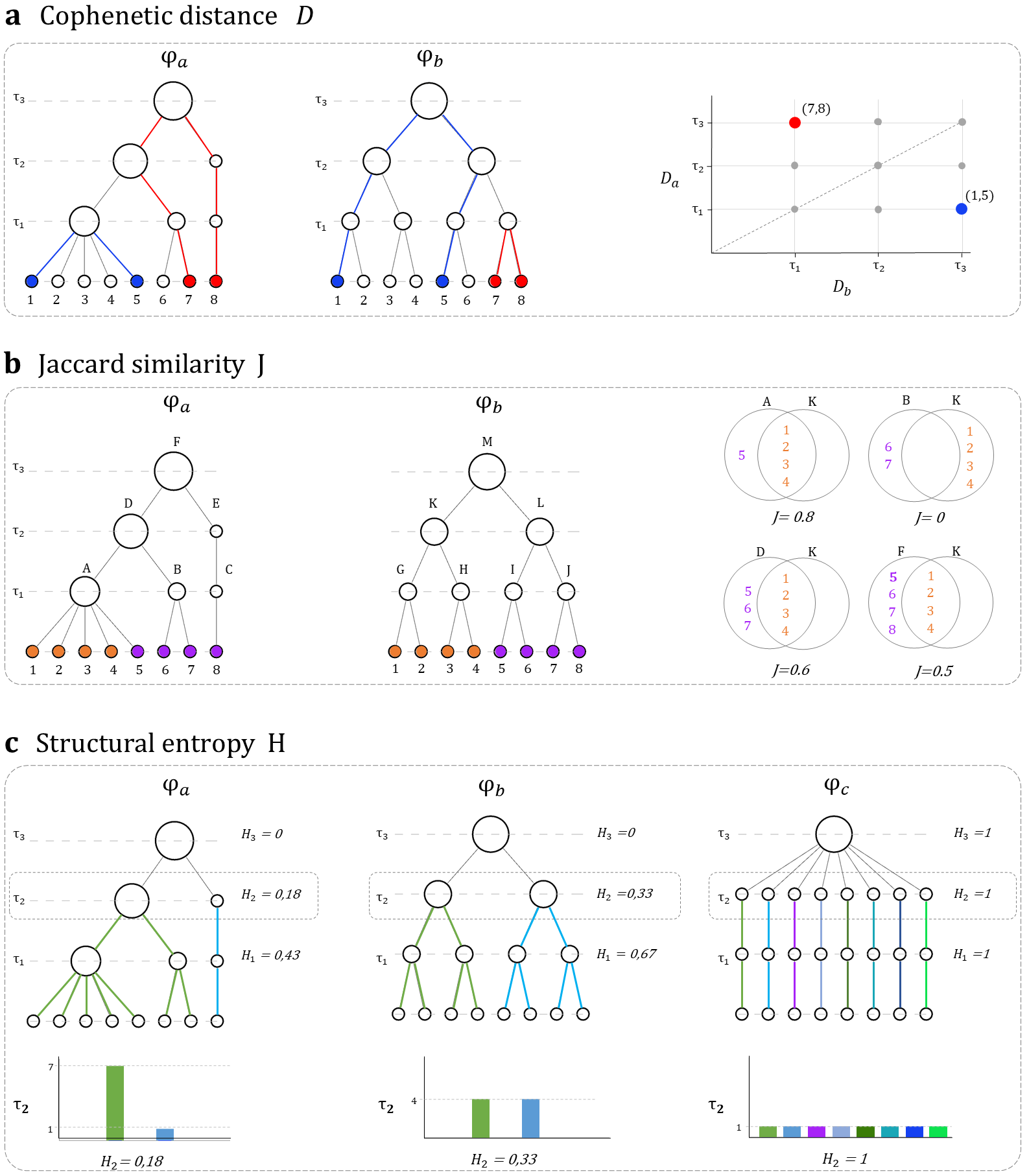}}
\caption{\textit{\textbf{Methods for hierarchical comparison in dendrogram toy models.} a) The hierarchical distance of clusters in dendrograms by computing $D$. b) The membership similarity between clusters by calculating $J$. c) The clustering diversity of dendrograms by measuring the $H_{\tau}$ of the topological structure. }}
\label{methods}
\end{figure}

To investigate the potential for scalar mismatch, we use three methods (Figure \ref{methods}) to compare the dendrograms (represented by $\varphi$) in terms of their members and structure.

\textbf{Cophenetic distance:} The Cophenetic distance $(D)$ allows us to compare the intensity of the relationships between the elements of the dendrogram.  It is defined as the level or threshold at which two terminal clusters $n_i$ and $n_j$ are joined together in the same cluster \citep{Sokal1962}. The larger the value connecting two clusters, the more distant they are from each other, and therefore the lower the potential for interaction between them.

In Figure \ref{methods}a we can see in $\varphi_a$ that terminal clusters $n_{1}$ and $n_{5}$ are first connected in threshold $\tau_1$, hence $D(n_{1},n_{5})=1$. For the same pair in $\varphi_b$, $D(n_{1},n_{5})=3$. Then $n_{1}$ and $n_{5}$ are more strongly connected in dendrogram $\varphi_a$. By plotting the distances of every city to all other cities within both dendrograms we are able to track down the dissimilar tendencies in the structuring process of these systems.

\textbf{Jaccard Similarity:} To compare the dendrograms, we use the Jaccard similarity index $(J)$ \citep{Jaccard}, defined as follows 

\begin{equation}\label{eq:(2)}
 J(c_{a},c_{b})=  \frac{c_{a} \cap c_{b}}{c_{a} \cup c_{b}},
\end{equation} 
where  $c_{a}$ is a cluster in $\varphi_a$ and $c_{b}$ is a cluster in $\varphi_b$, $c_{a} \cap c_{b}$ is the number of common terminal leaves $n_i \in N$ to both clusters, and $c_{a} \cup c_{b}$ is the total number of terminal leaves $n_i \in N$ in the two clusters.

This index calculates the proportion of elements that are common to two clusters, out of the total number of members in both.  In this paper, we compute $J$ between all clusters from both dendrograms, based on their common terminal elements in $N$. This allows us to investigate how cities cluster together and whether they follow the same organisational logic in both systems. The higher the $J$ value, the greater the similarity in the membership of the clusters. In Figure \ref{methods}b, we observe that the higher membership similarity between clusters in $\varphi_a$ and in $\varphi_b$ is given between $A \in \varphi_a$ and $K \in \varphi_b$, with $J=0.8$.

\textbf{Structural entropy:} This measure allows us to explore the agglomeration preferences between urban centres and to identify monocentric, polycentric, or disperse clustering processes across scales. To compute it, we look at the topology of the dendrogram, which is defined as the branching pattern relationship among its elements \citep{Phipps1971}. We are interested in exploring the organisational heterogeneity of the topology given by the diversity of the furcation patterns in the tree. We achieve this, by computing at each threshold $\tau$ the structural entropy $H_{\tau}$, which is a measure derived from Shannon's entropy function \citep{Shannon1948}, that determines the uncertainty of a city to join an urban cluster at that scale, as follows 

\begin{equation}\label{eq:(2)}
 H_{\tau}=  -\frac{1}{\log |C_{\tau}|}\sum_{ k \in \tau}  p_{k}  \log p_{k}.
\end{equation} 
$p_{k}$ is the probability defined as $p_{k} = \frac{|c_{k}|}{|N|}$, where $|c_{k}|$ is the cardinality (or size) of the cluster $c_k$ given by the number of elements $n_i \in N$ it has, and $|N|$ is the total number of elements $n_i \in N$ in the whole system. $|C_{\tau}|$ is the total number of clusters $c_{k}$ at $\tau$. The use of $\log |C_{\tau}|$ normalises the outputs at every threshold allowing the comparison across scales and between dendrograms. In Figure \ref{methods}c we can see the value of $H_{\tau}$ for every dendrogram at every threshold. For example, at $\tau=2$ the lowest value of $H_{\tau}$  is for $\varphi_a$ and the highest for $\varphi_c$. 

\section{Results and discussion}

In this section, we apply the methodology to Chile. This is an interesting case study, because it lies along an elongated territory, favouring longitudinal relations between cities, but possessing a contradicting centralised structure in terms of its administrative governance. 
The analysis highlights the mismatches between the imposed administrative structure and the natural organisation of the country emerging from the proxies for interactions.

\subsection{Scales of organisation}

\begin{figure}[h!]
\centering
\makebox[\textwidth][c]{\includegraphics[width=1\textwidth]{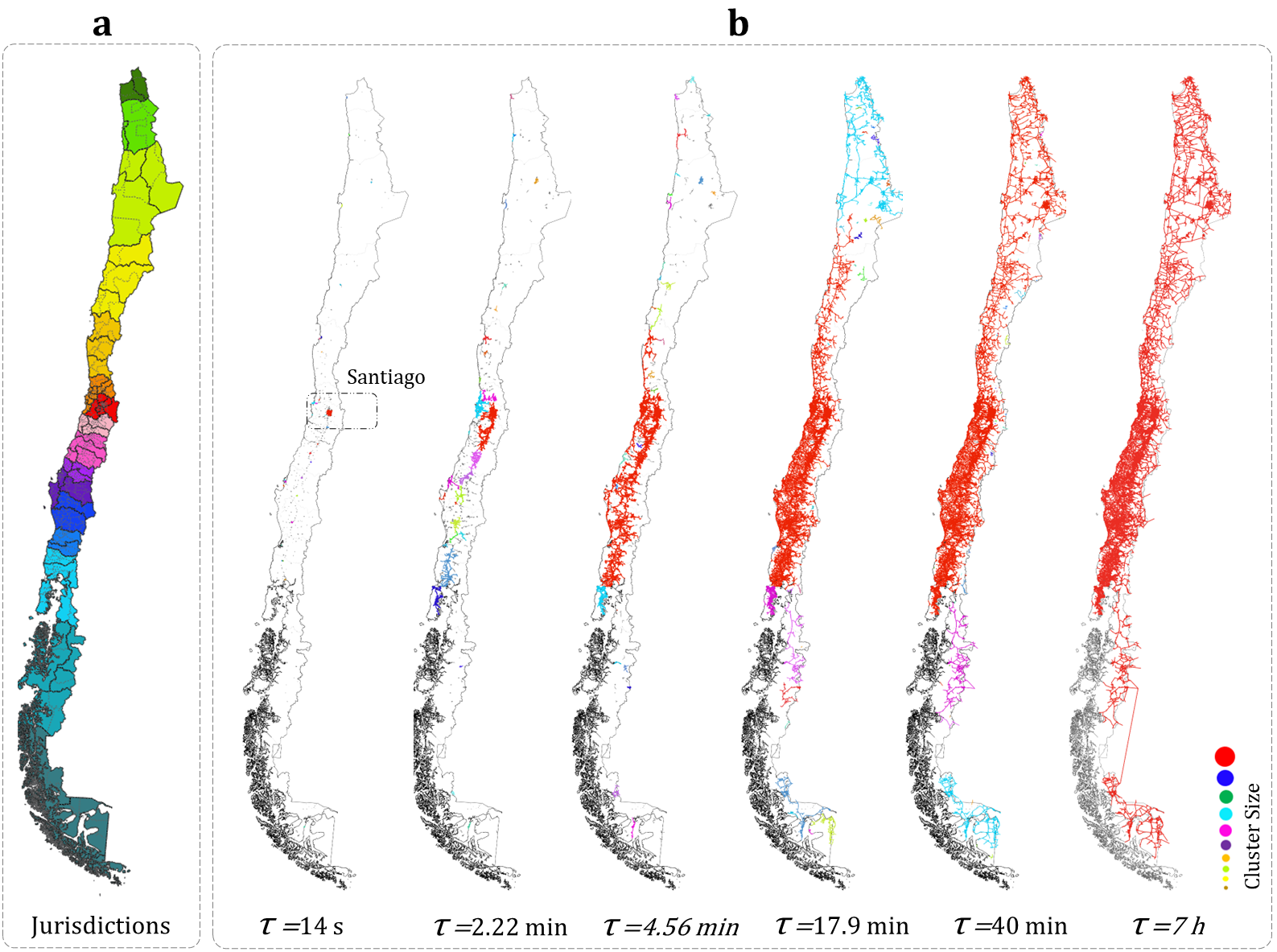}}
\caption{\textit{\textbf{Chilean administrative and urban connectivity organisation.} a) Limits of the administrative jurisdictions, where the colours show the different regions and the lines display the province limits. b) Sample of clusters at relevant percolation thresholds, from  the most local scale, corresponding to cities and towns, to more regional scales, until reaching the entire national cluster. At $\tau = 14$s for example, clusters are formed by street intersections that are on average 14 seconds apart, equivalent to 120 meters (an average block length) at a driving speed of 30 km/h. Connectivity at the city-scale level is detected at this threshold. The following threshold of 2.22min is equivalent to 2.8km road segments length between intersections travelled at 100km/h, which corresponds to the connectivity of clusters at intraregional scales.  Colours represent the rank of the clusters' size in terms of street network intersection count.}}
\label{maps}
\end{figure}

Percolation outputs allow us to visualise the formation of different systems of cities and regions along the thresholding process (Figure ~\ref{maps}b). Going from lower to higher time thresholds, that is from clusters with stronger to weaker ties, we observe that one of the first transitions generated at $\tau=14$s is in good correspondence with the urban extent of cities and towns. From now on the clusters arising at this threshold are considered the lowest scale of urban organisation for this analysis, and correspond to the terminal nodes $n_i \in N$ in the dendrogram. It is important to note that at this transition, the cluster sizes are evenly distributed along the national territory. As the threshold increases, we observe a consistent giant central cluster, which absorbs the neighbouring clusters in a radial manner. At $\tau=2.22$min, we see that this corresponds to a central urban corridor containing the capital city Santiago. 
At $\tau=4.56$min, this giant central cluster is now merged with almost all central urban areas, while other smaller clusters are still dispersed in the north and south without constituting agglomerations at regional scale. At the transition $\tau=40$min, the northern area is finally merged into a region and appended to the rest of the country. At the end of the percolation process at $\tau=7$h, the southern and Patagonian clusters are connected to the system, forming one cluster that covers all the national territory.

The process of percolation unveils an accentuated centralised tendency of cluster formation, with the pervasiveness of the capital city in the evolution of the clusters accompanied by a less prominent conformation of other sub-national structures. As evidenced by the different phase transitions, the system grows radially, from the centre where the capital is located, to the peripheral regions of the country. The predominance of the capital city in the spatial organisation of the national system is in good agreement with its political and administrative hegemony. Santiago not only accommodates almost half of the population, concentrates big part of the labour productivity and is responsible for 69\% of the GDP growth of the country \citep{OECD2017}, but it also concentrates most of the political and administrative power. 

\begin{figure}[h!]
\centering
\makebox[\textwidth][c]{\includegraphics[width=1\textwidth]{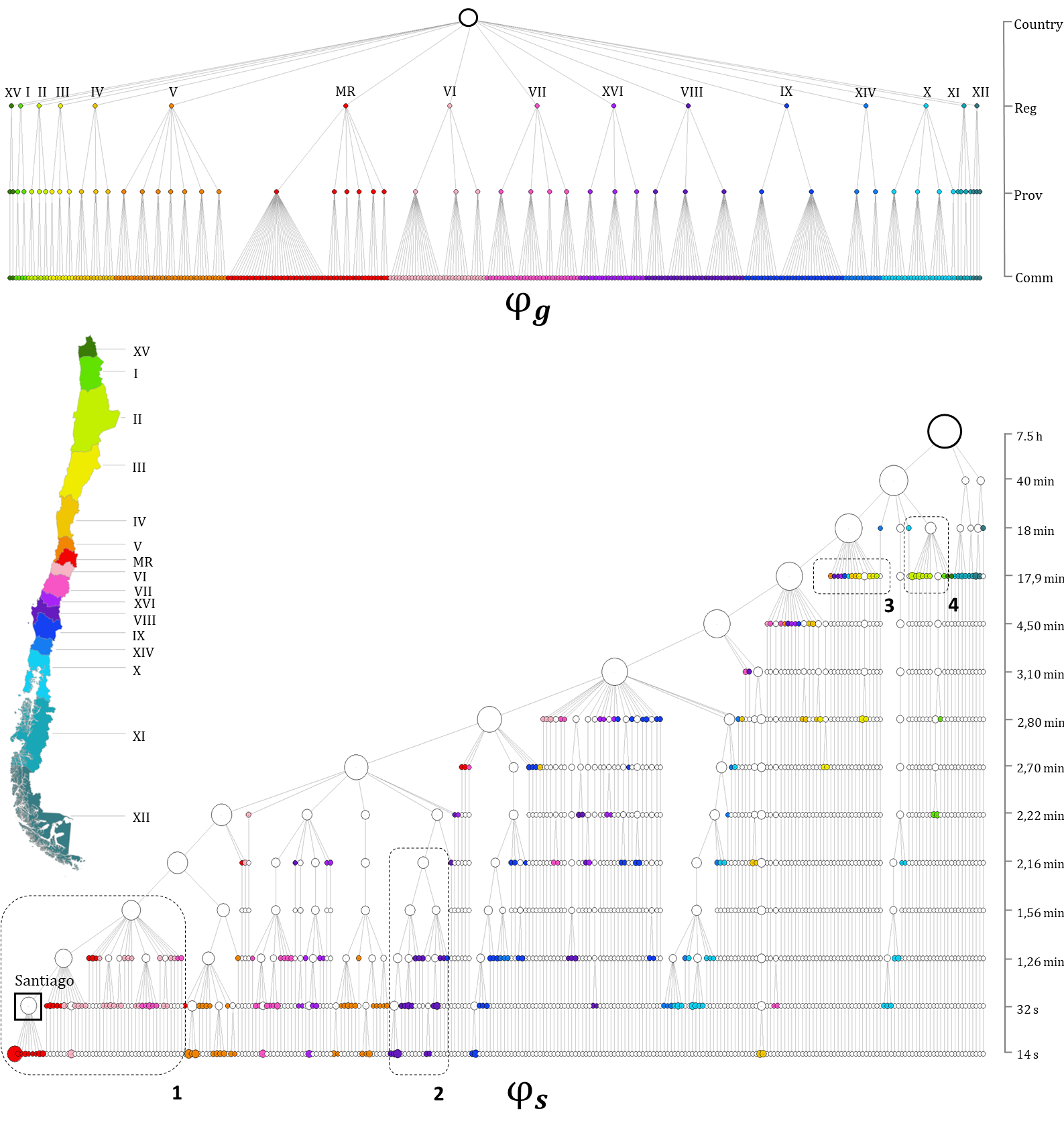}}
\caption{\textit{\textbf{Dendrograms representing the administrative and the urban hierarchical organisation of Chile.} $\varphi_g$ corresponds to the governance dendrogram extracted from the politico-administrative subdivision of the country, with communal, provincial, regional and national levels. $\varphi_s$ corresponds to the dendrogram of urban clusters resulting from the percolation process on the street network. Both dendrograms start with the same terminal clusters $n_i \in N$ representing cities and towns, and end with the national level at the top. Colour code in both corresponds to the administrative regional membership.}}
\label{dendros}
\end{figure}

\subsection{Regional formation}

Regions are made up of elements that are connected with varying levels of intensity and are distributed unevenly in space. While some regions are spatially cohesive as expected given their administrative boundaries, others are composed of isolated and more dispersed city systems. In this section, we compute the hierarchical distance between cities, to illustrate the diversity of intra-regional relationships. 

First, we can identify different patterns of formation of city systems by looking at the dendrograms in Figure \ref{dendros}. The urban connectivity patterns in $\varphi_s$ differ greatly from the national jurisdictional structure in $\varphi_g$, which does not show major configurational variation between regions. In the latter, the levels of hierarchy represent administrative scales (communes, provinces and regions), while in $\varphi_s$, the levels represent scales of organisation based on intensities of physical connectivity. In Figure \ref{dendros}, for example, we can see that in $\varphi_s:2$, a group of cities from the same region are clustered together at early thresholds, indicating lower hierarchical distance between them. In $\varphi_s:4$, a group of cities from the same region clusters together at the very end of the percolation process, indicating higher hierarchical distances between them and a weak potential for connectivity with the rest of the system. In $\varphi_s:3$, we identify cities that later in the thresholding process are still isolated, failing to connect with other cities before being absorbed by the larger cluster.

\begin{figure}[h!]
\centering
\makebox[\textwidth][c]{\includegraphics[width=1\textwidth]{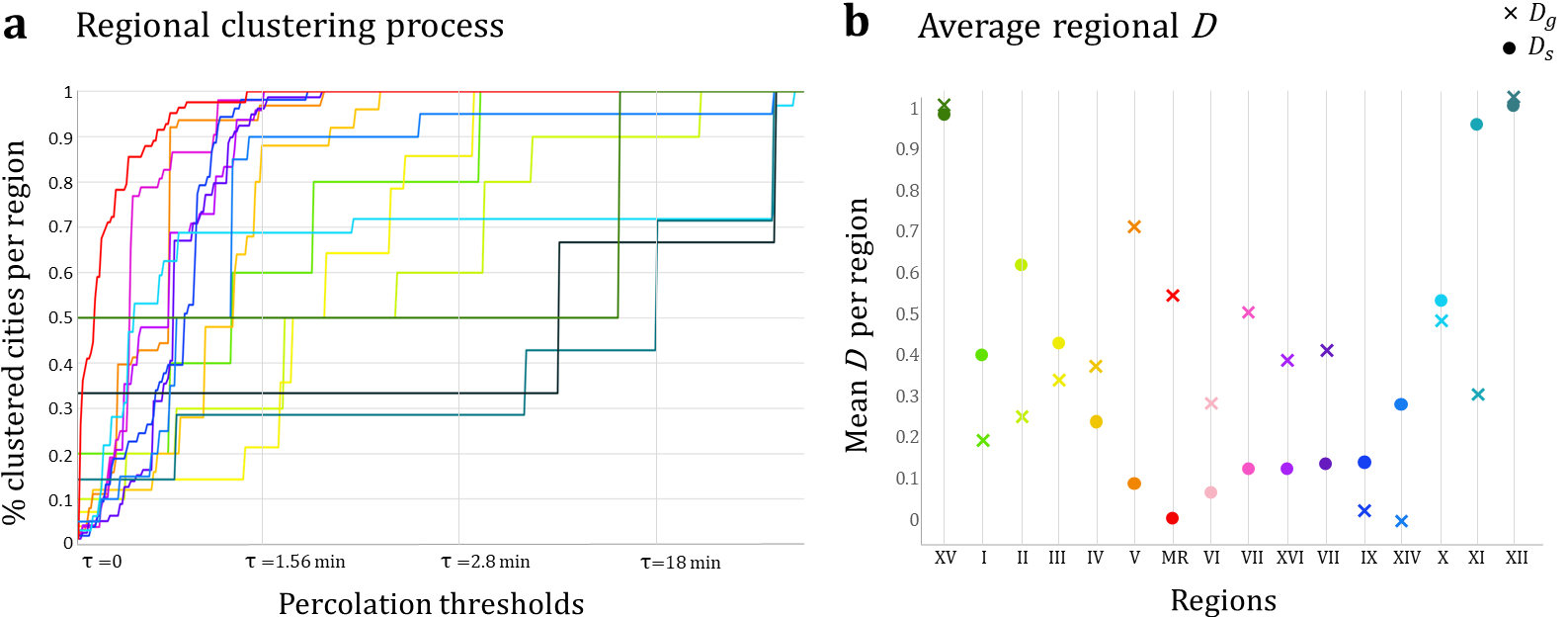}}
\caption{\textit{\textbf{Regional clustering process and hierarchical distance $D$.} In a) the percentage of clustered urban entities along the percolation process per region. In b) the mean regional cophenetic distances for the governance dendrogram $\varphi_g$ ($D_g$) and the urban connectivity dendrogram $\varphi_s$ ($D_s$). Central regions (V,MR,VI,VII) clustered faster during the percolation process in $a$ and have lower mean $D$ between regional elements in $b$.}}
\label{cdmean}
\end{figure}

We can measure the hierarchical distances between cities within each region using the cophenetic distance $D$. In Figure \ref{cdmean}b, we compare the average regional cophenetic distance of the urban connectivity dendrogram $\varphi_s$ ($D_s$) with the cophenetic distance from the governance dendrogram $\varphi_g$ ($D_g$). We see that the central regions have a lower $D_s$ average, while the extreme regions have higher values. In the case of $D_g$, the results do not show a clear pattern. The distances between cities in this case depend on the number of cities per region and not much on the inner hierarchical structure. 

\begin{figure}[h!]
\centering
\makebox[\textwidth][c]{\includegraphics[width=1.1\textwidth]{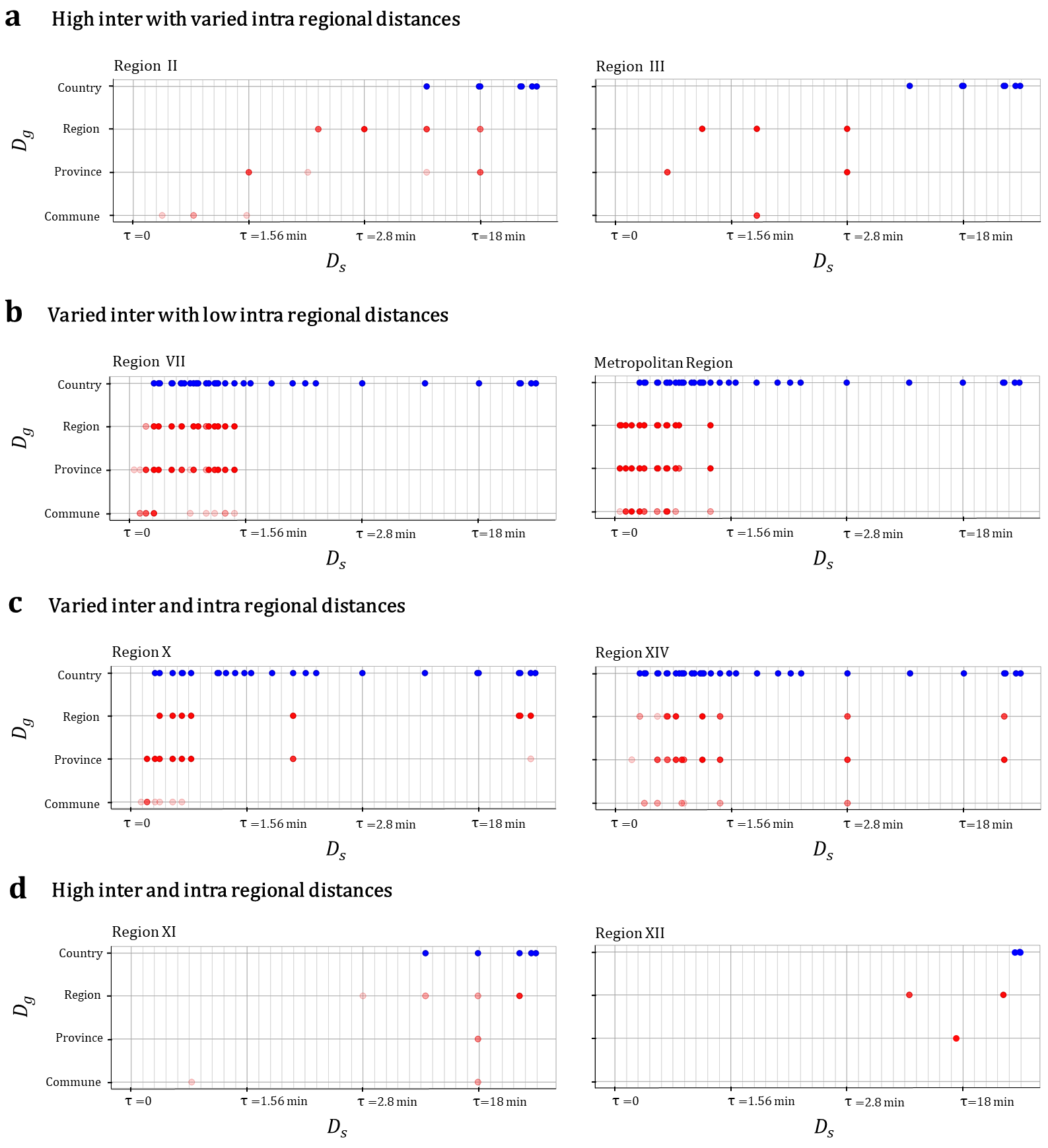}}
\caption{\textit{\textbf{Regional cophenetic distances ($D$) between dendrograms $\varphi_g$ and $\varphi_s$.} Each plot represents a single administrative region, displaying the results of the cophenetic distances from every city within the region to other cities in the same or different regions. Intraregional distances are shown in red and interregional distances in blue. The y-axis corresponds to the cophenetic distance in dendrogram $\varphi_g$ ($D_g$) with the different governance levels, and the x-axis corresponds to the  cophenetic distance in dendrogram $\varphi_s$ ($D_s$) with the thresholds from the street network percolation. The analysis allows us to identify various regional typologies which are grouped in a, b, c, and d. }}
\label{cds}
\end{figure}

In Figure \ref{cds}, we plot the cophenetic distances between elements in the urban connectivity dendrogram $\varphi_s$ ($D_s$) on the x-axis, and the distances in the governance dendrogram $\varphi_g$ ($D_g$) on the y-axis. Each of the plots extracts the distances between cities per regions. The red points highlight distances up to the regional level, which correspond to distances between cities within the same region. The blue points show distances within the national level, corresponding to inter-regional relationships. For example, Figure \ref{cds}a shows that in region II and III the lower $D_s$ distances are found between cities in the same communes, then between provinces and regions, following a natural order in both dendrograms. In these regions, the highest connectivity intensities are found between cities belonging to the same region (in red), and there is a much greater distance to the rest of the national system (in blue). In these cases, intraregional distances (blue) are very similar and overlap. In regions VII and XIII, intra-regional and inter-regional $D_s$ distances are similar, regardless of the governance distances $D_g$. This indicates that the regional system has a high level of connectivity with most of the cities in the system. In regions X and XIV, the  $D_s$ distances between the elements in dendrogram $\varphi_s$ do not represent the same organisation as those in $\varphi_g$. Some distances between cities within the same region are much higher than inter-regional distances. The intensity of connectivity in regions is not homogeneous, and in these cases some cities are less accessible to the connectivity of the regional system. Finally in regions XI and XII, inter- and intra- regional distances are both high, meaning low regional and national spatial connectivity.

The results of our analysis show the diverse scalar patterns of regional connectivity. In some regions, the hierarchical distances between the cities are shorter (Figure \ref{cds}b), which suggests a greater potential for spatial cohesion and alignment with the governance structure. This can facilitate communication and collaboration and make decision-making more efficient.  However, in regions with higher hierarchical distances (Figure \ref{cds}a and d), some cities may have less potential for regional interaction, and less access to resources and opportunities, which can hinder their development and lead to disparities within the region. Regions with uneven connectivity and fragmented city systems may struggle to develop a shared regional vision and goals, as is the case of the regions in Figure \ref{cds}c, making it difficult for decision makers to address regional challenges and promote the development of the region as a whole.

\subsection{Regional membership}

The output of this section reveals that potential strong relationships between urban systems are not always determined by politico-administrative definitions and are rather driven by the dynamics of the urban systems themselves. The multiplicity of regional membership within clusters is measured using the Jaccard similarity $J$, which provides insights into the mismatch between governance and the urban inner structural organisation.

\begin{figure}[h!]
\centering
\makebox[\textwidth][c]{\includegraphics[width=1\textwidth]{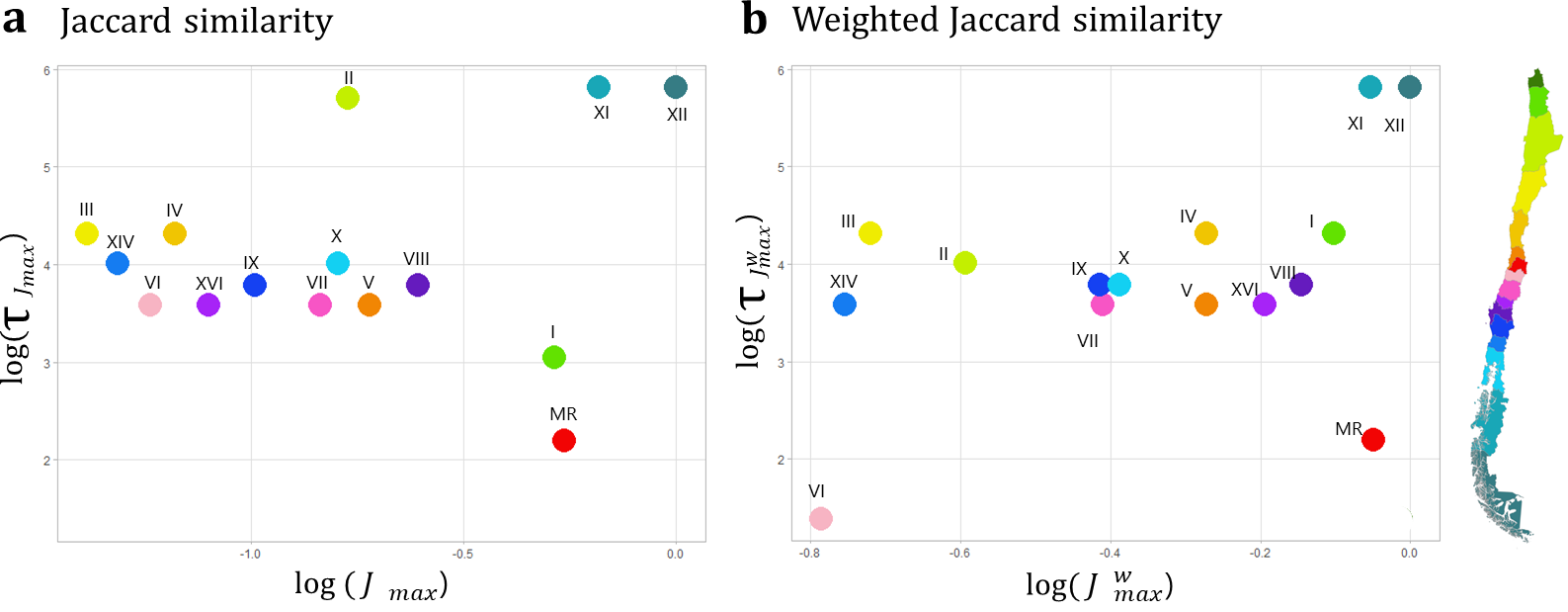}}
\caption{\textit{\textbf{Regional Jaccard Similarity between all clusters from dendrograms $\varphi_g$ and $\varphi_s$.} In the x-axis the $J$ maximum values per region and in the y-axis the percolation threshold from $\varphi_s$ at which the maximum $J$ is obtained. $J$ is computed in a) considering the number of cities and towns within clusters, and in b) using as a weight the size of the clusters given their street intersections count.}}
\label{jaccard}
\end{figure}

By examining the dendrograms in Figure \ref{dendros}, we can see that in $\varphi_s$:1, there is a cluster formed by cities from three different regions. In this case, cities from regions VI (light pink) and VII (pink) are first connected to cities in the Metropolitan Region (red) before connecting to other cities in their assigned regions. This is not the case in $\varphi_g$, where all cities belonging to the same region are first clustered together. To better understand the patterns observed in the dendrograms, we quantify $J$ between all clusters from both dendrograms in Figure \ref{dendros}. This allows us to identify in which regions systems of cities cluster in more or less accordance with their territorial delimitations. Figure \ref{jaccard} shows the value of the highest Jaccard similarity  $J$ for each region and the threshold at which it is obtained.

The results in Figure \ref{jaccard} also show that most of the regional clusters have their highest $J$ values at similar thresholds. This suggests that between these thresholds, we can detect regional formations in the urban system. In $a$, we use the number of terminal clusters $n_i \in N$ (without considering their size) to calculate the $J$ value. In $b$, we use the number of intersections in the street network within the clusters as a weight. As a result, larger cities will have a greater weight than small towns. This change leads to greater differences in $J$, providing more detailed information about potential mismatches.

As shown in Figure \ref{jaccard}, the metropolitan region $MR$, where the country's capital is located, has a high Jaccard similarity value at a very low threshold, indicating a good correspondence between the administrative organisation and the urban organisation resulting from the local interactions. On the other hand, regions $XI$ and $XII$ only reach high $J$ values at high thresholds. Cities in these regions are grouped following the governance regional membership, nevertheless, unlike the Metropolitan Region $MR$, these cities connect spatially very late in the percolation process. This can be explained by the fact that these regions are located in the extreme south of the country, wherecities are spatially segregated due to the fragmented geography. In contrast to the previous examples, Region $II$ has a very low Jaccard similarity value. This indicates that the cities belonging to this region do not tend to group into the same clusters. When examining Figure \ref{dendros}-$\varphi_s$, we observe that the cities from Region II, light green colour, are not contained in the same mother cluster. This indicates a large mismatch between territorial governance and the urban system.

The results indicate varying degrees of mismatch between the organisation of regional memberships. This information allows us to better understand how decisions made at one level of governance can align with the relationships of urban systems beyond jurisdictional boundaries. This can improve our understanding of the effects of policies on targeted urban systems and potential impacts on larger scales. When the scales of governance and the systems being governed match, it is more likely that measures can be effectively implemented. Otherwise, there may be a need for greater coordination between regional and lower levels of administration.

\subsection{Regional structure}

\makeatletter
\newenvironment{chapquote}[2][2em]
  {\setlength{\@tempdima}{#1}%
   \def\chapquote@author{#2}%
   \parshape 1 \@tempdima \dimexpr\textwidth-2\@tempdima\relax%
   \itshape}
  {\par\normalfont\hfill--\ \chapquote@author\hspace*{\@tempdima}\par\bigskip}
\makeatother

\begin{chapquote}{National Commission for Administrative Reform (CONARA) 1974}
``In the new regions a new homogeneous and equivalent institutionality will be established''
\end{chapquote}

\begin{figure}[h!]
\centering
\makebox[\textwidth][c]{\includegraphics[width=1\textwidth]{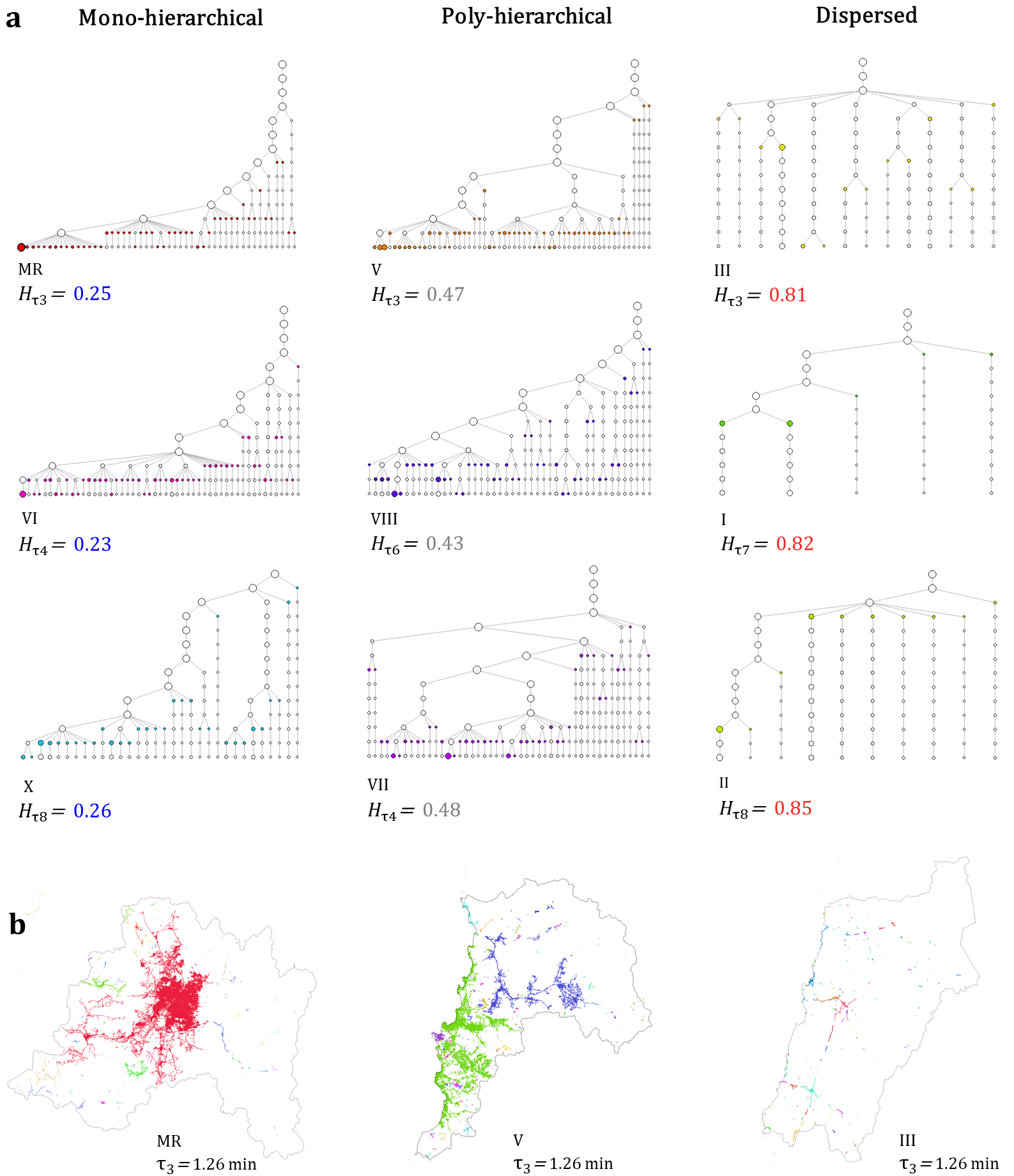}}
\caption{\textit{\textbf{Regional hierarchical organisations and structural entropy $H$.} In a) the sub-dendrograms from the different regions. The first column identifies mono-hierarchical regional structures, in the second column the poly-hierarchical regions, and in the last column the dispersed hierarchical organisations. The $H$ are computed for the thresholds with the maximum $J$. In b) the plots from three different regions in the relevant thresholds.}}
\label{Reg}
\end{figure}

From the national dendrogram we extracted a sub-dendrogram for each administrative region, keeping the same thresholds (Figure \ref{Reg}). In this case, the hierarchical comparison is made between all the regional dendrograms in order to visualise and quantify the heterogeneity of their topological organisation. This allows us to contrast the logic of intra-regional structures that emerge from the urban connectivity patterns with the monocentric administrative approach, that is replicated uniformly across regions in the country. Results show that although there are some regions that follow the mono-hierarchical organisation based on a regional capital, there are other poly-hierarchical as well as dispersed regional organisations. 

Figure \ref{Reg} shows the hierarchical dendrograms constructed for every region, where clusters of cities and towns $n_i \in N$ belonging to the same regional membership sit at the first scale of organisation. In mono-centric systems, as is the case of regions MR,VI and X, the percolation process evince a centre of growth that tends to assimilate single or scarcely clustered cities as the region develops. In polycentric structures, such as V,VIII and VII, two or more principal urban clusters develop in parallel, while in dispersed structures, see III, I and II, most cities and towns remain in isolation without forming systems of cities between regional constituent elements. In addition to comparing the topological differences between the dendrograms in Figure \ref{Reg}, we can also quantify these differences by measuring the structural entropy $H_{\tau}$. This can provide a more objective way to classifying the structures. For each of the thresholds in each dendrogram we measure the diversity of clustering patterns. The values in Figure \ref{Reg} correspond to the $H_{\tau}$ in thresholds at which each region obtained its highest value of $J$. In the case, when for a given threshold, a cluster contains most of the terminal clusters $n_i \in N$, the system will exhibit a skewed probability distribution, indicating that if a city or town is taken at random, it will most likely belong to that cluster. In information-theoretic terms, a lower entropy value $H_{\tau}$ indicates reduced uncertainty in clustering patterns. On the other hand, if the probability distribution of the $n_i \in N$ across clusters in a threshold is similar, the uncertainty in the clustering pattern will be higher and therefore the entropy $H_{\tau}$ will also be higher. When aggregation patterns are equiprobable, the entropy will be at its maximum.

For example, in Figure \ref{Reg}a, first column, we can see that in the dendrogram of the Metropolitan Region (MR) the system evolves in relation to the cluster that is initially defined by the country's capital city. As the clustering process progresses, clusters of cities and towns are attached to this main cluster. At different phase transitions, the dominance of the largest cluster and the absence of other significant clustering formations reflect the monocentric structure, which is consistent with the regional governance structure. The predominance of the cluster that contains the capital, reduces the uncertainty of the system, and therefore $H_{\tau}$ is also low, with $0.25$ in this case. In contrast, in regions like Valparaiso (V), in the second column, the clustering process evolves in the form of two independent intermediate systems before a regional system is established. This polycentric structure is a result of the low connectivity between coastal and valley cities, which is hindered by the Coastal Mountain Chain that tends to isolate both sides. The regional capital also named Valparaiso, is part of the cluster formed in the coastal side. In this case, there is more uncertainty about how cities are grouped, as there is no clear dominant cluster. Instead, different systems of cities are forming. The entropy of poly-hierarchical dendrograms is higher than that of mono-hierarchical structures, with $H_{\tau}=0.47$ in Valparaiso. Finally, dendrograms in the last column show a more disperse furcation structure, where small clusters of cities are scattered during the whole process. An example is the Atacama Region (III), where the thresholding process does not show phase transitions defining clusters formed by systems of cities, and rather cities remain isolated until merging with the largest cluster at the latest levels. In this case, the lack of dominant clustering leads to a higher entropy value of $H_{\tau}$ = 0.81.

The variety of regional hierarchical structures could suggest the need for a more diverse system of governance that can take into account local particularities. The current regional administrative approach centred around a well-defined development hub or capital city, may not always align with the spatial organisation of regions, leading to a mismatch that can have negative effects on planning. The outputs of the analysis show some regions with a mono-hierarchical organisation which aligns well with the administrative logic, while others do not. Regions with more diverse and dispersed hierarchical structures may benefit from a more flexible, multiscalar governance approach that takes into account the varying capacities of different cities to collaborate and influence development. This could help promote positive spillover effects and support the overall growth of the region.

\section{Conclusions}

In this paper, we employ several quantitative methods, including cophenetic distance, Jaccard similarity, and structural entropy, to compare dendrograms from different regions. This enables us to investigate the mismatches between the physical urban connectivity and the geographical administrative structures. 
We begin by examining the hierarchical distance between elements, calculating the cophenetic distance. When there are mismatches, we observe regions where cities are widely distant in terms of their relational intensity, which can hinder their ability to function as a unified entity. In such cases, governance faces greater challenges in establishing common development goals and promoting equitable growth. Next, we examine the membership of the dendrograms by computing the Jaccard similarity. When mismatches are present, we may find cases where the administrative boundaries of regions do not align with the clustering of cities. This can lead to cities from different regions being grouped together with higher levels of connectivity rather than forming strong regional clusters first. This is important for governance, in order to understand the impact of regional policies at different levels.
Finally, we analyse the regional organisational structure by measuring the structural entropy of the dendrograms. In cases where mismatches exist, the clustering process may not align with the administrative logic based on a single centre of regional development, resulting in polycentric rather than monocentric regional structures. This presents a major challenge for governance, as it highlights the need to include different regional development strategies, including polycentric structures.

In this work, we focused on the scalar composition of regions in Chile, so that territorial disparities can be addressed. We found that current regional management approaches often overlook local dynamics and the heterogeneities within city systems. Our analysis provides valuable insights into the hierarchical structure of urban systems in different regions, and emphasises the importance of considering the scalar composition of regions in decision-making. We found that scalar mismatches suggest that a one-size-fits-all approach to regional governance may not be effective, and that decision-makers should take into account the specific organisation of each region in order to promote the development of the region as a whole. In future work, we plan to expand our analysis to include functional networks, bringing together structure and flows.

Overall, this work contributes to advancing our understanding of the complexity of the spatial regional morphology from a scalar approach. The different methods for comparing dendrograms are crucial for a comprehensive understanding of scalar mismatch, providing a framework for studying the interplay between the administrative organisation and the structure that emerges from the spatial connectivity of the urban system.


\bibliographystyle{agsm}

\section*{Acknowledgements}

VM thanks The National Research and Development Agency (ANID) for the financial support provided as a PhD scholarship Becas Chile.

\section*{Author contributions statement}
VM developed the theoretical framework, designed the experiments and conducted the analysis. CM and EA contributed to the theoretical framework . VM wrote the paper, and all authors contributed to the structure and fine tuning of the manuscript.

\section*{Competing interests}

The authors declare no competing interests.

\end{document}